\documentclass[floats,floatfix,showpacs,preprint,prd,superscriptaddress,onecolumn,nofootinbib]{revtex4}
\usepackage{graphicx, epsfig, amssymb} %include figure files
\usepackage{epsfig}
\usepackage{amsmath, amsfonts}
\usepackage{bm} %include bold math: \bm{} creates bold letters in math mode
\usepackage{color}
\usepackage[usenames]{xcolor}
\usepackage[colorlinks=true,linkcolor=blue,urlcolor=blue,filecolor=black,citecolor=red,pdfstartview=FitV,pdftitle={},pdfsubject={},pdfkeywords={},pdfpagemode=None,bookmarksopen=true]{hyperref}
\usepackage{float}
\usepackage{pgfplots}
\pgfplotsset{compat=1.12}
\newcommand{\be}{\begin{equation}}
\newcommand{\ee}{\end{equation}}
\newcommand{\bea}{\begin{eqnarray}}
\newcommand{\eea}{\end{eqnarray}}

\def \dr {\partial_r}

\begin{document}

%%%%%%%

\title{Axionic black branes with conformal coupling}

%%%%%%%

\author{Adolfo Cisterna}
\email{adolfo.cisterna@ucentral.cl}
\affiliation{Universidad Central de Chile, Vicerrector\'ia acad\'emica, Toesca 1783, Santiago, Chile}
\affiliation{Dipartimento di Fisica, Universit\`a di Trento,Via Sommarive 14, 38123 Povo (TN), Italy}

\author{Cristi\'an Erices}
\email{crerices@central.ntua.gr}
\affiliation{Department of Physics, National Technical University of Athens, Zografou Campus, GR 15773, Athens, Greece}
\affiliation{Universidad Cat\'olica del Maule,
Av. San Miguel 3605, Talca, Chile.}

\author{Xiao-Mei Kuang}
\email{xmeikuang@gmail.com}
\affiliation{Center for Gravitation and Cosmology, College of Physical \\ Science and Technology, Yangzhou University, Yangzhou 225009, China}

\author{Massimiliano Rinaldi}
\email{massimiliano.rinaldi@unitn.it}
\affiliation{Dipartimento di Fisica, Universit\`a di Trento,Via Sommarive 14, 38123 Povo (TN), Italy}
\affiliation{TIFPA - INFN, Via Sommarive 14, 38123 Povo (TN), Italy}

\begin{abstract}
We find neutral and charged black branes solutions with axion fields in the context of a conformally coupled gravitational theory in four dimensions.
These solutions describe AdS black branes supported by axion fields that break translational invariance at the boundary, providing for momentum dissipation.
The conformally coupled scalar field is regular inside and outside the event horizon and there is no need of any self-interaction, obtaining in this way solutions without fine-tuned parameters. We analyze the thermodynamics of our solutions considering the effects of the axion charges and it is shown that axionic and electric charges must be related such that the conformal scalar field does not contribute to the mass. We compute the holographic DC conductivity and we show how it is affected by the inclusion of the conformal scalar field, which provides a temperature independent behavior. We include a non-linear axionic contribution given by a k-essence term that modifies the DC conductivity providing for more general behaviors. Finally, we endorse our solutions with rotation showing that angular momentum is sustained by the axion charges.

\vskip -15pt

\noindent
\end{abstract}
%%%%%%

\maketitle

%%%%%%

\section{Introduction}

For any well-posed gravitational theory such as Einstein general relativity (GR) or any of its extensions, the existence of black hole solutions is a matter of primary interest. Black holes offer the perfect arena to study the theory in the strong gravity regime providing information about the causal structure of the spacetime and astrophysical relevant predictions. Moreover, their semiclassical descriptions through the study of their thermodynamical properties provide the perfect setup to study quantum gravity effects that would lead to the characterization of a fundamental theory of quantum gravity \cite{Hawking:1974sw}.
However, in standard GR, to find black hole solutions dressed with matter fields is typically a non-trivial task due to the existence of the topological censorship theorem \cite{Friedman:1993ty} as well as the no-hair conjecture \cite{Bekenstein:1998aw}.

On one hand, topological obstructions in four dimensions can be evaded by relaxing the asymptotic behavior through the inclusion of a cosmological constant, obtaining static solutions with planar and hyperbolic horizons \cite{Birmingham:1998nr} along with several solutions with non-trivial topology at infinity \cite{Vanzo:1997gw}\footnote{Solutions with non-trivial topology at spatial infinity are obtained compactifying the base manifold directions. They are usually dubbed topological black holes.}. In contrast, topological obstructions are weaker in dimension $D>4$ allowing asymptotically flat solutions with non-spherical topology such as black rings \cite{Emparan:2001wn} and diverse black object solutions \cite{Horowitz:2012nnc}. Moreover, for higher dimensional GR, the unicity theorems \cite{{Carter1,Israel1,Wald1}} are no longer valid permitting black hole solutions such as the Schwarzschild-Tangherlini black hole \cite{Tangherlini:1963bw} and the black $p$-brane.

On the other hand, the no-hair conjecture states that black holes can not be described by any different quantity apart from its mass, electric charge and angular momentum \cite{Ruffini:1971bza}. This implies that, after gravitational collapse, black holes can be only characterized by quantities that follows a Gauss type law, namely charges that can be measured at infinity. Any other characteristic of the matter that falls into the black hole is lost. This is because the assumptions forbid black holes with non-trivial matter fields apart from the electromagnetic one, which is the well known Kerr-Newman solution \cite{Kerr:1963ud}. As shown in \cite{Janis:1968zz}, there are no non-trivial regular solutions in GR when minimally coupled scalar fields are considered. The no-hair conjecture renders the scalar field trivial and the solution is nothing else than the Schwarzschild black hole. Nevertheless, this conjecture can be circumvented by allowing the introduction of suitable potentials or nonminimal couplings to the matter field. The nonminimal coupling of conformal type with electromagnetic interaction and in absence of cosmological constant, was considered first by Bronnikov, Melnikov and Bocharova \cite{Bocharova:1970skc} and Bekenstein \cite{Bekenstein:1974sf,Bekenstein:1975ts}\footnote{Another way to circumvent the conjecture is to consider complex scalar fields that do not share the same symmetries of spacetime, in such a way that its contribution to the strees-energy tensor it does. Using this approach Kerr black holes with scalar hair have been numerically constructed in \cite{Herdeiro:2014goa}.}. This is the first counter-example to the no-hair conjecture using scalar fields and it represents a black hole only in four dimensions \cite{Xanthopoulos:1992fm}. Due to the fact that the scalar field does not introduces any new integration constant on the backreaction these solutions are dubbed ``solutions with secondary hair". 
However, although the metric of this so called ``BMBB" black hole turns out to be the extreme Reissner-Nordtr\"{o}m (RN) solution, the scalar field diverges at the horizon making its physical properties difficult to interpret. As it was shown in \cite{Martinez:2002ru,Martinez:2005di}, this physical pathology can be fixed by introducing a cosmological constant pushing the scalar field singularity behind the event horizon. This solution, dubbed the ``MTZ" black hole, possesses a spherical or hyperbolic horizon depending on the sign of the cosmological constant and exists only for a particular combination between the cosmological and the quartic self-interacting coupling constant\footnote{A number of solutions were found including more general self-interactions in four and in higher dimensions \cite{Herdeiro:2015waa}.}. No planar solution is allowed to exists.
Precisely, planar solutions are typically affected by singular behaviors, symptom of a shortage of a curvature scale on the horizon. In \cite{Bardoux:2012aw} the authors, based on a family of metrics that accomplish a weaker version of the Birkhoff's theorem, have constructed four dimensional asymptotically AdS black holes with planar horizons supported by matter represented by $p$-forms\footnote{Applications of these ideas have provided several new regular black brane solutions \cite{Bardoux:2012tr,Caldarelli:2013gqa,Bardoux:2013swa,Caldarelli:2016nni}.}. Then, regular solutions can be found by charging the horizon with homogeneously distributed axionic charges along planar directions. These axion fields endow spacetime with an effective intrinsic curvature scale making possible to regularize black hole solutions. In \cite{Bardoux:2012tr} the authors have constructed a non-trivial planar version of the MTZ black hole including, apart from the MTZ solution ingredients, two axion fields given by three forms originated from two Kalb-Raimond potentials. In here, not only the the scalar fields is nonminimally coupled to gravity, but also the axion fields. These solutions were also generalized to the case in which the nonminimal coupling is arbitrary and to the case of higher dimensions \cite{Caldarelli:2013gqa}.

It is known that the flat geometry of the horizon opens the possibility to study holographic applications based on the gauge/gravity duality. The AdS/CFT correspondence \cite{Maldacena:1997re} establishes a duality between gravitational theories in $D$ dimensions and conformal field theories in the $(D-1)$-dimensional boundary. In this scenario, black holes became suitable laboratories to study strongly coupled systems opening a new range of applicability of gravitational solutions at the service of condense matter systems \cite{Gubser:2002tv,Witten:1998qj}. In this respect, planar/toroidal black holes with scalar fields possess special relevance due, in particular, to their applications in the dual description of superconductor systems \cite{Hartnoll:2009sz,Horowitz:2010gk}. Hairy black holes can undergo spontaneous undressing up in a phase transition process reminiscent of nonzero condensate behavior in unconventional superconductors. To successfully describe real materials using the holographic tools \cite{Gubser:2002tv,Witten:1998qj}, it is very important to include a mechanism of momentum dissipation. Several methods are employed in order to accomplish this, such as the so-called scalar lattice technique implemented by a periodic scalar source \cite{Q1,Q2}, the framework of massive gravity where the diffeomorphism invariance of the theory breaks down in the bulk \cite{M1,M2,M3,Kuang:2017rpx}, or the Q-lattice model in which the phase of a complex scalar fields breaks the translational invariance of the theory \cite{q1,q2}. A very simple way to describe systems with momentum dissipation is the case of massless scalar fields that depends linearly on the base manifold coordinates. This technique was introduced in \cite{andrade} as an effective way to break translational invariance on the dual field theory obtaining several novel properties for the dual condense matter sector. Several solutions within these ideas were reported in \cite{Davison,Gout,Baggioli,Alberte,Bagg,Kuang,Cisterna:2017jmv}.\footnote{A very interesting application is the construction of homogeneous black string and black p-branes with negative cosmological constant, with no more ingredients that minimally coupled scalar fields \cite{Cisterna:2017qrb}. }\\
This paper is devoted to the construction of AdS black brane solutions with a conformally coupled scalar field where the translational invariance at the boundary is broken by means of axion fields that depend linearly on the base manifold directions. This solution represents the generalization to the conformal coupling case of \cite{andrade} in four dimensions and can be viewed as an economic way to obtain the planar version of the BMBB solution.\\
The paper is organized as follows: Sec. II presents the theory under consideration. In Sec. III the solutions are exposed and their principal features are discussed. Sec. IV provides a general description of the black brane thermodynamic analysis while in Sec. V we described the holographic DC conductivity and Hall angle for these solutions. We include a non-linear k-essence term in order to modify the classical expected behavior of conductivities on these types of models. Finally we conclude in Sec. VI.

\section{The theory}
We consider four dimensional gravity with a cosmological constant interacting with a matter source given by a conformally coupled real scalar field and two free axions, described by the action
\begin{equation}
S[g,\phi,\psi_I]=\int{d^4x\sqrt{-g}\left[\kappa(R-2\Lambda)-\frac{1}{2}(\partial\phi)^2-\frac{1}{12}\phi^2 R-\frac{1}{2}{\sum_{I=1}^2}(\partial\psi_I)^2\right]},   \label{action}
\end{equation}
where $\kappa\equiv\frac{1}{16\pi G}$ and $G$ is the four dimensional Newton's constant. The field equations are
\begin{eqnarray}
\kappa(G_{\mu\nu}+\Lambda g_{\mu\nu})&=&\frac{1}{2} T^{\phi}_{\mu\nu}+\frac{1}{2}T^{\psi}_{\mu\nu},\label{eqgrav}\\
\left(\Box-\frac{1}{6}R\right)\phi&=&0,\label{eqphi}\\
\Box\psi_I&=&0\label{eqaxion},
\end{eqnarray}
where $\Box\equiv g^{\mu\nu}\nabla_{\mu}\nabla_{\nu}$. The energy-momentum tensor is given by contributions from the scalar and axion field which are respectively
\begin{eqnarray}
T^{\phi}_{\mu\nu}&=&\partial_\mu\phi\partial_\nu\phi-\frac{1}{2}g_{\mu\nu}(\partial\phi)^2+\frac{1}{6}(g_{\mu\nu}\Box-\nabla_\mu \nabla_\nu+G_{\mu\nu})\phi^2,\\
T^{\psi}_{\mu\nu}&=&\sum_{I=1}^{2}\left(\partial_\mu\psi_I\partial_\nu\psi_I-\frac{1}{2}g_{\mu\nu}(\partial\psi_I)^2\right).
\end{eqnarray}
We look for static and planar four dimensional metrics, whose expression are given by
\begin{equation}
ds^2=-F(r)dt^2+\frac{dr^2}{F(r)}+r^2(dx^2+dy^2),  \label{metric}
\end{equation}
where $0\leq r<\infty$, $0\leq x\leq\beta_x$ and $0\leq y\leq\beta_y$. The base manifold is assumed to be compact, without boundary and of vanishing curvature, i.e. it is locally isometric to flat space $\mathbb{R}^2$.
Imposing the axion fields to depend only on the boundary directions the Klein-Gordon equation for each axion field is trivially integrated yielding
\begin{eqnarray}\label{axiongral}
\psi_I=\zeta_{Ii} x^i+\alpha_I,
\end{eqnarray}
where $x^1\equiv x$, $x^2\equiv y$ and $\zeta_{Ii}, \alpha_I$ are integration constants. Due to the form of the axion kinetic term in the action, the axion field enjoys global ISO(2) symmetry. As it was pointed out in \cite{Caldarelli:2016nni}, this means that for planar solutions of the field equations, the global ISO(2) symmetry is isomorphic to the spatial isometries of the conformal boundary. Nevertheless, solution \eqref{axiongral} breaks completely the global ISO(2) symmetry. As a matter of choice, we are interested in solutions that breaks translation symmetry in the conformal boundary, for which $\alpha_I=0$. This is because, from an holographic point of view, it is more interesting, since it induces momentum dissipation in the dual field theory. We have preserved only the $SO(2)$ symmetry of the conformal boundary, allowing to rearrange the expression for the axion field in terms only of two integration constants $\lambda_I$, which translate into the constraint $\sum_{I=1}^{2}(\zeta_{Ii}\zeta_{Ij}-\lambda_{I}^2\delta_{ij})=0$. In this way, we may write the solution as $\psi_I=\lambda x_{I}$.

%$\lambda^2=\frac{1}{2}\sum_{i=1}^{2}\sum_{I=1}^{2}\zeta_{Ii}\zeta_{Ii}$

\section{Four dimensional black brane solution}
The field equations \eqref{eqgrav}, \eqref{eqphi} and \eqref{eqaxion} admit an exact solution where the metric, scalar and axion fields are given by
\begin{eqnarray}
ds^2&=&-\frac{(r-3\lambda l)(r+\lambda l)^3}{r^2 l^2}dt^2+\frac{r^2 l^2}{(r-3\lambda l)(r+\lambda l)^3}dr^2+r^2(dx^2+dy^2), \label{solu}\\
\phi&=&2\sqrt{3}\frac{\lambda l}{r+\lambda l},\label{scalarfield}\\
\psi_I&=&2\sqrt{3}\lambda x_I\label{axionfield},
\end{eqnarray}
where we have redefined the axion parameter $\lambda\rightarrow2\sqrt{3}\lambda$, set $\kappa=1$ for simplicity and defined the AdS radius  $l^{-2}:=-\frac{\Lambda}{3}$. The metric \eqref{solu} describes a planar black hole, provided the cosmological constant is negative, with an asymptotically AdS behavior as it can be seen by the large $r$ behavior of the components $g_{tt}\sim -\frac{r^2}{l^2}+\mathcal{O}(r^0)$, $g_{rr}\sim \frac{l^2}{r^2}+\mathcal{O}(r^0)$. There is a curvature singularity at the origin as it can be checked by evaluating the Ricci scalar
\begin{equation}
R=-\frac{12}{l^2}+\frac{12 \lambda^2}{r^2},
\end{equation}
which is dressed by a single event horizon located at $r_+=3\lambda l$, when $\lambda$ is positive, and at $r_+=-\lambda l$ when $\lambda$ is negative. In the latter case the scalar field diverges at the event horizon resembling what occurs with the BMBB configuration, making its physical interpretation not clear as the entropy is not well defined. However, unlike the BMBB and MTZ configurations, the scalar field has no poles when $\lambda$ is positive, being regular everywhere and hereafter, we analyze this well-behaved solution. In contrast with the case of the MTZ black hole configuration, where the cosmological constant is fine-tuned by the quartic self-interacting coupling constant, here the cosmological constant is completely arbitrary. This means that this solution naturally emerges as the most economic way to obtain a toroidal black hole with a conformally coupled scalar field which is free of singularities in four dimensions. The inclusion of both free axion fields allows us to obtain planar solutions and regularizes everywhere the conformally coupled scalar field in a simpler way respect to the case with conformally coupled scalar field with quartic self-interaction.

An electrically and magnetically charged black hole is obtained by adding the Maxwell term,
\begin{equation}\label{actionMax}
-\frac{1}{4}\int d^4 x\sqrt{-g}F^{\mu\nu}F_{\mu\nu},
\end{equation}
to the action (\ref{action}), obtaining the same metric \eqref{solu} and axion field \eqref{axionfield}, but with a scalar field and gauge potential of the form
\begin{align}
\phi&=\frac{\sqrt{Q_E^2+Q_M^2+12\lambda^4 l^2}}{\lambda(r+\lambda l)},\ &\ A&=-\frac{Q_E}{r}dt+\frac{Q_M}{2}(xdy-ydx). \label{sgfields}
\end{align}
At large $r$, scalar field is approximated by
\begin{equation}
\phi=\frac{\phi_1}{r}+\frac{\phi_2}{r^2}+\mathcal{O}(r^{-3}),
\end{equation}
where
\begin{align}\label{defphis}\phi_1&\equiv\lambda^{-1}\sqrt{Q_E^2+Q_M^2+12l^2\lambda^4},&\phi_2&\equiv-l \sqrt{Q_E^2+Q_M^2+12 l^2\lambda^4}.
\end{align}
It is worth to point out that, as it was shown in \cite{HMTZ, Hertog}, the boundary conditions that ensures the AdS symmetry of the scalar field asymptotic behavior are $\{\phi_1=0,\phi_2\neq0\}$, $\{\phi_1\neq0,\phi_2=0\}$ and $\phi_1^2=\alpha\phi_2$, where $\alpha$ is a constant without variation. However, only the third boundary condition allows a non-vanishing scalar field. By means of \eqref{defphis}, and in order to satisfy this boundary condition, a relation between the parameters of the solution must be fulfilled as follows,
\begin{align}\label{AdScriterium}
\lambda^4&=\frac{Q_E^2+Q_M^2}{(\alpha^{2}-12)l^2},&|\alpha|&>2\sqrt{3}.
\end{align}
It can be checked that the dominant energy condition is satisfied for the charged and neutral black hole. In fact, by computing the stress-energy tensor in the orthonormal frame,
\begin{equation}
T^{ab}={\rm diag}\left(\frac{12 r^2-6l^2 \lambda^4}{r^4},-\frac{12 r^2-6l^2 \lambda^4}{r^4},-\frac{6l^2\lambda^4}{r^4},-\frac{6l^2\lambda^4}{r^4}\right),
\end{equation}
we note that its canonical form is of type I according to the classification in \cite{Hawking:1973uf}. We can identify the energy density $\rho$ and the principal pressures $p_a$ ($a=1,2,3$), as
\begin{align}
\rho&=-p_1=\frac{12 r^2-6l^2 \lambda^4}{r^4},&p_2&=p_3=-\frac{6l^2\lambda^4}{r^4},
\end{align}
verifying directly that $\rho\geq0$ and $-\rho\leq p_a\leq\rho$ for $r\geq\lambda l$.  Note that the event horizon $r_+>3\lambda l$ covers this region, implying that the dominant energy condition holds in the causally connected region of the spacetime.\\
Our solutions are easily extended to the case in which they possess angular momentum. Due to our planar base manifold this can be made by applying an improper gauge transformation on the time coordinate and on one of the transverse manifold directions upon identification of this last coordinate. We explicitly show this construction and its thermodynamic analysis in the Appendix A.
\section{Black hole thermodynamics} \label{secthermo}

For a more complete understanding of the black hole solutions above we perform a complete thermodynamical analysis using the Euclidean approach. In this context, the partition function for a thermodynamical ensemble is identified with the Euclidean path integral in the saddle point approximation around the classical Euclidean solution \cite{GH}. \\
We consider spacetimes with a manifold of topology $\mathbb{R}^2\times\mathbb{R}^2$ where the first plane $\mathbb{R}^2$ is parametrized by the radial coordinate $r$ and the Euclidean time $\tau$, while the second refers to the base manifold assumed to be compact of volume $\sigma$ and, as we mentioned, spanned by coordinates $x$ and $y$. Therefore the Euclidean continuation of the black hole metric \eqref{metric} reads,
\begin{equation}
ds_E^2=N^2(r)F(r)d\tau^2+\frac{dr^2}{F(r)}+r^2(dx^2+dy^2),
\end{equation}
and the scalar, axionic and gauge field are
\begin{align}
\phi&=\phi(r),\ &\psi_{I}&=\psi_{I}(x^{i}),\ &A&=A_{\tau}(r)d\tau+A_x(y)dx+A_y(x)dy,\
\end{align}
where $x^{1}=x$ and $x^{2}=y$. These plane coordinates range as $0\leq\tau\leq\beta$, $r_+\leq r<\infty$, $0\leq x\leq\beta_x$ and $0\leq y\leq\beta_y$ where the period $\beta$ is identified with the inverse temperature and the Euclidean action is related to the Gibbs $\mathcal{G}$ free energy by $I_E=\beta \mathcal{G}$. These considerations lead to the reduced Euclidean action $I_E$ to take the Hamiltonian form,
\begin{equation}\label{reduced_action}
I_E=\beta\int_{r_+}^{\infty}dr\int_{0}^{\beta_x}dx\int_{0}^{\beta_y}dy(N\mathcal{H}-A_{\tau}\mathcal{G})+B_E
\end{equation}
with $B_E$ a surface term. The reduced constraints are given by
\begin{equation}
\begin{aligned}
\mathcal{H}=&\frac{r^2}{8\pi G}\left[\left(1-\frac{4\pi G}{3}\phi^2\right)\left(\frac{\dr F}{r}+\frac{F}{r^2}\right)+\Lambda\right]+\frac{r^2}{6}\left[F(\dr\phi)^2-\left(\dr F+\frac{4F}{r}\right)\phi\dr\phi-2F\phi\dr^2\phi\right]\\
&+\frac{1}{2}\sum_{i=1}^2(\partial_{x^i}\psi_{I})^2+\frac{1}{2r^2}(\partial_x A_y-\partial_y A_x)^2+\frac{1}{2r^2}(\pi^r)^2,\\
\mathcal{G}=&\dr \pi^r,
\end{aligned}
\end{equation}
where $\pi^r$ stands for the electromagnetic field momentum defined by,
\begin{equation}
\pi^r=-\frac{r^2 A'_{\tau}}{N}.
\end{equation}
This is the only non-vanishing momenta conjugated to the fields.
In order to get a well-defined variational problem, the Euclidean action must be a differential functional of the canonical variables $\{F,A_x,A_y,\phi,\pi^r,\psi_I\}$, i.e. $\delta I_E=0$ on-shell. This means that the boundary term $B_E$ must cancel all the boundary terms induced by the variations of the bulk term when the total variation of the action is performed. It is easy to check that the equation of motion obtained by varying the reduced action with respect to $\{N,F,A_{\tau},A_x,A_y,\phi,\pi^r\}$ are consistent with the original Einstein equations. It turns out that $N$ is a constant which, without loss of generality, can be taken to be $N=1$. It is certainly evident, from \eqref{reduced_action}, that $N$ and $A_{\tau}$ are Lagrange multipliers whose associated constraints are the Hamiltonian one $\mathcal{H}=0$ and the Gauss law $\mathcal{G}=0$ for the electromagnetic field conjugate momentum. The former, in addition to variations respect to $F$ and $\phi$, provides a closed set of equations whose solution is given by the expressions for $F$ and $\phi$ found in \eqref{metric}. The later determines a Coulomb form for the electric potential. Maybe the less direct equations of motion are given by variations respect to $\psi_{I}$ and $A_{x^i}$ which turn out to be respectively,
\begin{align}
\partial_{x^i}^2\psi_{I}&=0,&\partial_{x^i}F_{xy}=0,
\end{align}
whose solution, after a suitable redefinition of the constants, provides a linear dependence on the coordinates for the fields in agreement with the solution found previously (\ref{axiongral}) and (\ref{sgfields}). Finally, the variation respect to the momenta $\pi^r$ determines an equation trivially satisfied by its definition.

The variational principle on the boundary term gives
\begin{equation}\label{deltaBE}
\begin{aligned}
\delta B_E=&\beta\sigma\left[A_{\tau} \delta\pi^r-\frac{r N}{8\pi G}\left(1-\frac{4\pi G}{3}\phi^2-\frac{4\pi G}{3}r\phi\phi'\right)\delta F-\frac{r^2 N}{6}\left(4 F \phi'+F'\phi\right)\delta\phi+\frac{r^2 N}{3}F\phi\delta\phi'\right]^{\infty}_{r_+}\\
&-\int_{r_+}^{\infty}dr\frac{N}{r^2}\left\{\left[\int_{0}^{\beta_x}dx(\partial_y A_x-\partial_x A_y)\delta A_x\right]^{y=\beta_y}_{y=0}-\left[\int_{0}^{\beta_y}dy(\partial_y A_x-\partial_x A_y)\delta A_y\right]^{x=\beta_x}_{x=0}\right\}\\
&-\int_{r_+}^{\infty}dr\left\{\left[\int_{0}^{\beta_y}dy\partial_x\psi_{1}\delta\psi_{1}\right]^{x=\beta_x}_{x=0}+\left[\int_{0}^{\beta_x}dx\partial_y\psi_{2}\delta\psi_{2}\right]^{y=\beta_y}_{y=0}\right\},
\end{aligned}
\end{equation}
where we have used the fact that the volume of the base manifold is $\sigma=\beta_x\beta_y$. Hereafter $N$ is considered as a constant chosen to be $N=1$. From the last two boundary terms the contribution of the topological and axionic charge of the system can be identified. These terms lead to the variation of the magnetic charge multiplied by the magnetic potential as well as the variation of the axionic charge multiplied by its associated chemical potential.

Requiring regularity of the metric at the horizon yields $\beta F'(r_+)=4\pi$ giving the value for the temperature,
\begin{equation}
T=\frac{16}{9\pi}\frac{\lambda}{l}. \label{temp}
\end{equation}
The variation of the fields on the event horizon is given by
\begin{equation}\label{delta_rh1}
\begin{aligned}
\delta F\rvert_{r_+}&=-\frac{4\pi}{\beta}\delta r_+,&\delta\phi\rvert_{r_+}&=\delta\phi(r_+)-\phi'\rvert_{r_+}\delta r_+,
\end{aligned}
\end{equation}
\begin{equation}\label{delta_rh2}
\begin{aligned}
\delta\psi_{I}\rvert_{r_+}&=2\sqrt{3} x^{i} \delta \lambda,&\delta\pi^r\rvert_{r_+}&=\delta Q_E,&\delta A_x\rvert_{r_+}&=\frac{\delta Q_M}{2}y,&\delta A_y\rvert_{r_+}&=-\frac{\delta Q_M}{2}x.
\end{aligned}
\end{equation}
It is convenient to define the effective Newton constant at the horizon $\tilde{G}_+$ as \footnote{In here we have reestablished the Newton constant $G$ for clearness. Note that we set $1/16\pi G=1$ when evaluating our solution.},
\begin{equation}
\tilde{G}_+=\frac{G}{\left(1-\frac{4\pi G}{3}\phi(r_+)^2\right)}.
\end{equation}
With this definition and \eqref{deltaBE}, \eqref{delta_rh1} and \eqref{delta_rh2}, the variation of the boundary term at the horizon is
\begin{equation}\label{dBrh}
\delta{B_E(r_+)}=\delta\left(\frac{A_+}{4\tilde{G}_+}\right)+\beta\Phi_{e}\delta(\sigma Q_E)+\beta\Phi_M\delta(\sigma  Q_M)+\beta\Phi_{\psi_{1}}\delta(-2\sqrt{3}\sigma\lambda)+\beta\Phi_{\psi_{2}}\delta(-2\sqrt{3}\sigma\lambda),
\end{equation}
where $A_{+}=\sigma r_+^2$ is the horizon area. Additionally we have conveniently defined
\begin{align}
\Phi_E&\equiv \frac{Q_E}{r_+},&\Phi_M&\equiv \frac{Q_M}{r_+},&\Phi_{\psi_{1}}&\equiv 2\sqrt{3}\lambda r_+,&\Phi_{\psi_{2}}&\equiv 2\sqrt{3}\lambda r_+,
\end{align}
identifying the chemical potentials for the electric, magnetic and axion fields as $\Phi_E,\Phi_M,\Phi_{\psi_{I}}$, respectively. We reordered the variations in \eqref{dBrh} using the fact that $\sigma$ is fixed to make a clear contact with the conserved charges later. Hereafter, we work in the grand canonical ensemble, where the temperature $T=\beta^{-1}$ and the chemical potentials at the horizon are fixed. Then, by virtue of these boundary conditions at the horizon, the boundary term can be integrated there, giving
\begin{equation}
B_E(r_+)=\frac{A_+}{4\tilde{G}_+}+\beta \Phi_{e}(r_+) (\sigma Q_E)+\beta \Phi_M (\sigma  Q_M)+\beta\Phi_{\psi_{1}}(-2\sqrt{3}\sigma\lambda)+\beta\Phi_{\psi_{2}}(-2\sqrt{3}\sigma\lambda).
\end{equation}
The variation of fields at infinity are
\begin{equation}\label{delta_inf1}
\begin{aligned}
\delta F\rvert_{\infty}&=-12 \lambda\delta\lambda-\frac{24 l \lambda^2\delta\lambda}{r}-\frac{12l^2\lambda^3\delta\lambda}{r^2},&
\delta\phi\rvert_{\infty}&=\frac{\delta\phi_1}{r}+\frac{\delta\phi_2}{r^2}+\mathcal{O}(r^{-3}),
\end{aligned}
\end{equation}
\begin{equation}\label{delta_inf2}
\begin{aligned}
\delta\psi_{I}\rvert_{\infty}&=2\sqrt{3} x^{i} \delta \lambda,&\delta\pi^r\rvert_{\infty}&=\delta Q_E,&\delta A_x\rvert_{\infty}&=\frac{\delta Q_M}{2}y,&\delta A_y\rvert_{\infty}&=-\frac{\delta Q_M}{2}x.
\end{aligned}
\end{equation}

Then, from \eqref{deltaBE} and \eqref{delta_inf1}-\eqref{delta_inf2} we obtain an expression for the variation of the boundary term at infinity as follows,
\begin{equation}
\delta B_E(\infty)=\beta\sigma48 l \lambda^2 \delta\lambda+\frac{\beta\sigma}{3 l^2}(2\phi_2\delta\phi_1-\phi_1\delta\phi_2).
\end{equation}
The integrability condition $\delta^2 B_E(\infty)=0$ can be imposed implying that $\phi_2=\phi_2(\phi_1)$. Additionally, since $\beta$ is fixed the boundary term at infinity generically takes the form
\begin{equation}
B_E(\infty)=\beta\sigma16 l  \lambda^3+\frac{\beta\sigma}{3l^2}\int\left(2\phi_2-\phi_1\frac{d\phi_2}{d\phi_1}\right)d\phi_1.
\end{equation}
Having computed the boundary terms and since the value of the reduced action on-shell is the boundary term $B_E$, we obtain,
\begin{equation}\label{reducedaction}
\begin{aligned}
I_E=&B_E(\infty)-B_E(r_+)\\
=&\beta\sigma16 l  \lambda^3+\frac{\beta\sigma}{3l^2}\int\left(2\phi_2-\phi_1\frac{d\phi_2}{d\phi_1}\right)d\phi_1-\frac{A_+}{4\tilde{G}_+}-\beta \Phi_{e}(r_+) (\sigma Q_E)-\beta \Phi_M (\sigma  Q_M)\\&-\beta\Phi_{\psi_{1}}(-2\sqrt{3}\sigma\lambda)-\beta\Phi_{\psi_{2}}(-2\sqrt{3}\sigma\lambda),
\end{aligned}
\end{equation}
up to an arbitrary additive constant without variation.
Since the Gibbs free energy is related to the Euclidean action as $I_E=\beta \mathcal{G}=\beta \mathcal{M}-\mathcal{S}-\beta \Phi_E\mathcal{Q}_E-\beta \Phi_M\mathcal{Q}_M-\beta\Phi_{\psi_{1}}\mathcal{Q}_1-\beta\Phi_{\psi_{2}}\mathcal{Q}_2$ in the grand canonical ensemble, the mass $\mathcal{M}$, the electric charge $\mathcal{Q}_E$, the magnetic charge $\mathcal{Q}_M$, the axionic charges $\mathcal{Q}_i$ and entropy $\mathcal{S}$ are computed by means of the standard thermodynamical relations. From \eqref{reducedaction} is straightforward that, the mass generically is given by
\begin{equation}\label{gralmass}
\begin{aligned}
\mathcal{M}&=\left(\frac{\partial}{\partial \beta}-\beta^{-1}\Phi_E\frac{\partial}{\partial \Phi_E}-\beta^{-1}\Phi_M\frac{\partial}{\partial \Phi_M}-\beta^{-1}\Phi_{\psi_I}\frac{\partial}{\partial \Phi_{\psi_I}}\right)I_E\\
&=16\sigma l \lambda^3+\frac{\sigma}{3l^2}\int\left(2\phi_2-\phi_1\frac{d\phi_2}{d\phi_1}\right)d\phi_1,
\end{aligned}
\end{equation}
whereas the entropy $\mathcal{S}$, axionic charges $\mathcal{Q}_i$, electric charge $\mathcal{Q}_E$ and magnetic charge $\mathcal{Q}_M$ read
\begin{align}
\mathcal{S}&=\left(\beta\frac{\partial}{\partial \beta}-1\right)I_E=\frac{A_+}{4\tilde{G}_+},&\mathcal{Q}_i&=-\frac{1}{\beta}\frac{\partial I_E}{\partial\Phi_{\psi_I}}=-2\sqrt{3}\lambda\sigma,\\
\mathcal{Q}_E&=-\frac{1}{\beta}\frac{\partial I_E}{\partial \Phi_E}=\sigma Q_E,&\mathcal{Q}_M&=-\frac{1}{\beta}\frac{\partial I_E}{\partial \Phi_M}=\sigma Q_M.
\end{align}
Thus, it turns out from these results, that the first law of black hole thermodynamics is satisfied,
\begin{equation}
d\mathcal{M}=TdS+\Phi_E d\mathcal{Q}_E+\Phi_M d\mathcal{Q}_M+\Phi_{\psi_1} d\mathcal{Q}_1+\Phi_{\psi_2} d\mathcal{Q}_2,
\end{equation}
which was expected since this is a consequence of the fact that the Euclidean reduced action attains an extremum.
The positivity of the entropy requires $\tilde{G}_+>0$ providing a lower bound for the axion parameter,
\begin{equation}\label{entrocon}
\lambda^4>\frac{Q_E^2+Q_M^2}{180 l^2},
\end{equation}
which implies an upper bound for the constant $\alpha$ through \eqref{AdScriterium} such that $|\alpha|<8\sqrt{3}$.
Therefore, we see that, unlike the uncharged case, the axion parameter is restricted by the electric and magnetic charge otherwise there is an unphysical negative entropy.
The precise functional relation between the leading and subleading term of the scalar field in the asymptotic region, can be determined by demanding that the scalar field respect the asymptotic AdS invariance. As we are interested in holographic applications, we will consider \eqref{AdScriterium}, with which the boundary condition vanishes the integral term in \eqref{gralmass} giving,
\begin{equation}
\mathcal{M}=16\sigma l \lambda^3.
\end{equation}
Therefore, there is no scalar field contribution in the black hole mass. As it was pointed out in \cite{AAM}, this is because to preserve AdS symmetry of the scalar field in the asymptotic region requires a vanishing contribution of the scalar field to the mass.

The rotating solution can be referred in Appendix A, in which for simplicity we have analyzed the neutral case. The main property of the rotating black brane is that the finiteness condition on the conserved charges imposes a relation between the angular momentum and the axionic parameter. This means that the axion field can provide angular momentum to physically acceptable solutions whose mass $\mathcal{M}$ and angular momentum $\mathcal{J}$ are respectively given by
\begin{align}
\mathcal{M}&=8\sigma l \lambda(3\omega_0^2+2\lambda^2),&\mathcal{J}&=24\sigma l^2 \lambda\omega_0\sqrt{\omega_0^2+\lambda^2},
\end{align}
where
\begin{equation}
\omega=\frac{\omega_0}{\sqrt{\omega_0^2+\lambda^2}}\ ,
\end{equation}
with $\omega_0$ a constant without variation.

The local stability can be analyzed by computing the specific heat at fixed chemical potentials giving,
\begin{equation}
C_{\Omega_+, \Phi_{\psi_I}}=\frac{3\pi\sigma(r_+^2+9l^2\omega_0^2)^{3/2}(2 r_+^2+9l^2\omega_0^2)}{2 r_+(r_+^2+18l^2\omega_0^2)},
\end{equation}
which is positive always. In consequence, the black brane can always reach thermal equilibrium with a heat bath.

\section{Holographic DC conductivities and Hall angle}
Now we move on to study the DC conductivities $\sigma_{DC}$ and the Hall angle $\theta_H$ of the holographic theory dual to the charged hairy black hole. These transport properties have recently been widely studied in various theories because the conductivities do not evolve in
the radial direction and hence can be analytically obtained by calculating the values at the horizon. It was discussed in \cite{q2,Blake:2014yla} that $\sigma_{DC}$ usually contains two terms, i.e., $\sigma_{DC}=\sigma_{ccs}+\sigma_{diss}$, where $\sigma_{ccs}$ is the `charge-conjugation symmetric' part \cite{Charmousis:2010zz} while $\sigma_{diss}$ is related to the charge $Q_E$ of the black hole and it is divergent in a translationally invariant theory. The relation between the Hall angle and $\sigma_{diss}$ and the corresponding scaling were carefully studied in \cite{Blake:2014yla,Zhou:2015dha}.\\
Thus, we will use the techniques of \cite{q2} to study the features of $\sigma_{DC}$ and $\theta_H$ of our charged hairy black hole.
To proceed, we turn on only the following relevant perturbations because the remaining perturbations are decoupled and have no relevance to our study,
\begin{eqnarray}
&&\delta A_x=-E_x t+a_x,~~\delta A_y=-E_y t+a_y,\nonumber\\
&&\delta g_{tx}=r^2h_{tx},~~\delta g_{rx}=r^2h_{rx},~~\delta g_{ty}=r^2h_{ty},~~\delta g_{ry}=r^2h_{ry},\nonumber\\
&&\delta\psi_1=\Psi_1,~~\delta\psi_2=\Psi_2,
\end{eqnarray}
where $E_x$ and $E_y$ are constant, while $a_x,a_y,h_{tx},h_{ty},h_{rx},h_{ry},\Psi_1,\Psi_2$ are all functions of the radial coordinate $r$.\\
Then, the two perturbed Maxwell equations are
\begin{eqnarray}
&&F' a_x'+F a_x''+Q_E h_{\text{tx}}'+Q_M \left(F' h_{\text{ry}}+F h_{\text{ry}}'\right)=0,\label{eq-Per-Max1}\\
&&F' a_y'+F a_y''+Q_E h_{\text{ty}}'-Q_M \left(F' h_{\text{rx}}+F h_{\text{rx}}'\right)=0,\label{eq-Per-Max2}
\end{eqnarray}
where the prime denotes the derivative to $r$. From the above equations, we define two conserved currents\footnote{In principle, we can also define the current $r^2 F^{rx}$ because of Maxwell equation $(r^2 F^{rx})'=0$, however, we have defined $J_t=Q_E=-r^2 F^{rt}=r^2 (A_t)^{'}$ in the black hole solution \eqref{sgfields}. In order to be consistent, we chose $J_i=-r^2 F^{ri}(i=x,y)$.  }
\begin{eqnarray}
J_x=-r^2 F^{rx}=-F a_x'-Q_E h_{tx}-Q_M F h_{ry},\\
J_y=-r^2 F^{ry}=-F a_y'-Q_E h_{ty}+Q_M F h_{rx},
\end{eqnarray}
which satisfy $\frac{dJ_x(r)}{dr}=\frac{dJ_y(r)}{dr}=0$ due to the Maxwell equations \eqref{eq-Per-Max1} and \eqref{eq-Per-Max2}.
This implies that $J_x$ and $J_y$ do not depend on $r$. In addition, according to the AdS/CFT dictionary, the holographic DC conductivities are determined by the conserved currents in the asymptotic boundary. As mentioned above, since $J_x$ and $J_y$ are independent of $r$, we shall evaluate them at the horizon instead of on the boundary.\\
To impose the regularity conditions at the horizon, it is convenient to work in Eddington-Finklestein coordinates $(v, r)$ with $v=t-\int \frac{dr }{F}$. Thus, the regular conditions at the event horizon require the gauge field to take the form  \cite{Donos:2014cya}
\begin{equation}
\label{ReguA}a_x=-E_x \int \frac{dr }{F},~~a_y=-E_y \int \frac{dr }{F}\end{equation}
while the perturbed metric reads
\begin{equation}\label{ReguH}
h_{rx}=\frac{h_{tx}}{ F},~~h_{ry}=\frac{h_{ty}}{ F}.
\end{equation}
Moreover, we have $F (r_+)\sim 4\pi T(r-r_+)$ and we set $\Psi_{1,2}$ to be constant near the horizon \cite{q2}. Then, substituting the conditions \eqref{ReguA} and \eqref{ReguH} into the $rx$ and $ry$ components of the Einstein equation, we can  solve for  $h_{tx}(r_+)$ and $h_{ty}(r_+)$. Thus, the conductivities  can be obtained
\begin{eqnarray}
\sigma_{xx}&=&\frac{\partial J_x(r_+)}{\partial E_x}=\frac{3 \left(12 \lambda ^4+Q_M^2+Q_E^2\right) \left(36 \lambda ^4+3
   Q_M^2+Q_E^2\right)}{2 Q_E^2 \left(36 \lambda ^4+5 Q_M^2\right)+9 \left(12 \lambda
   ^4+Q_M^2\right){}^2+Q_E^4},\\
\sigma_{yy}&=&\frac{\partial J_y(r_+)}{\partial E_y}=\sigma_{xx},\\
\sigma_{xy}&=&\frac{\partial J_x(r_+)}{\partial E_y}=\frac{6 Q_E Q_M \left(12 \lambda ^4+Q_M^2+Q_E^2\right)}{2 Q_E^2 \left(36 \lambda ^4+5
   Q_M^2\right)+9 \left(12 \lambda ^4+Q_M^2\right){}^2+Q_E^4},\\
\sigma_{yx}&=&\frac{\partial J_y(r_+)}{\partial E_x}=-\frac{6 Q_E Q_M \left(12 \lambda ^4+Q_M^2+Q_E^2\right)}{2 Q_E^2 \left(36 \lambda ^4+5
   Q_M^2\right)+9 \left(12 \lambda ^4+Q_M^2\right){}^2+Q_E^4}=-\sigma_{xy},
\end{eqnarray}
where we have considered the event horizon $r_+=3\lambda$ and the temperature $T=16\lambda/9\pi$ in \eqref{temp} by setting $l=1$.
The DC conductivity and the Hall angle are
\begin{eqnarray}
\sigma_{DC}&:=&\sigma_{xx}(Q_M=0)=\frac{3 \left(12 \lambda ^4+Q_E^2\right)}{36 \lambda^4+Q_E^2},\\
\theta_H&:=&\frac{\sigma_{xy}}{\sigma_{xx}}=\frac{2 Q_E Q_M}{36 \lambda ^4+3 Q_M^2+Q_E^2},
\end{eqnarray}
and we see that $\sigma_{DC}$ and $\theta_H$ are finite.\\
We rewrite the DC conductivity as
\begin{equation}\label{conduc}
\sigma_{DC}=1+\frac{2Q_E^2}{36 \lambda ^4+Q_E^2}=1+\frac{2(\alpha^2-12)}{24+\alpha^2}=\frac{3\alpha^2}{24+\alpha^2},
\end{equation}
where we have used \eqref{AdScriterium} in the second equality.
When $Q_E=0$, we have $\sigma_{DC}=1$ which is consistent with the result for neutral black hole found in \cite{q2}. However, the  term $\frac{2Q_E^2}{36 \lambda ^4+Q_E^2}$ is very different from $\sigma_{diss}$ in the case without scalar field \cite{andrade}, which is divergent as $\lambda\to 0$. Theories in which  other scalar fields appear, other than the minimally coupled axions, have been considered for example in \cite{Jiang:2017imk}. We observe that, even in very exotic cases, in which other fields with nonminimal couplings are included, the conductivity still behaves as if those fields were not acting, reproducing the same result found in \cite{andrade}. Of course, once the theory under consideration modifies the location of the horizon in term of these new contributions to the system, the dependence on temperature will changes. Nevertheless the asymptotic behavior of the conductivity remains unchanged. In our case, we observe that the conformally coupled scalar field $\phi(r)$ modifies the backreaction of the black brane solution such that $\sigma_{DC}$ is temperature independent. Actually, after considering \eqref{AdScriterium}, we observe that it does not depend on $\lambda$ at all, and then by means of (\ref{temp}) there is no dependence on the temperature.
Thus, the holographic transport features in our model are more like that in neutral black hole whose DC conductivity is a finite constant in the dual theory.
%It is reasonable because the axion fields are minimally coupled as well as the Maxwell fields.
%This is explicit because in the solution \eqref{solu},\eqref{axionfield} and \eqref{sgfields}, the Maxwell field does not explicitly appear in the metric, so it behaves as a probe field.
We note that the DC conductivity in this model satisfies $\sigma_{DC}>1$ by considering \eqref{AdScriterium} into \eqref{conduc}. The only dependence that $\sigma_{DC}$ possesses is the dependence on the constant $\alpha$. From (\ref{AdScriterium}) we know that the absolute value of $\alpha$ must be strictly greater than $2\sqrt{3}$. On the other hand we observe that in order to reproduce solutions with positive entropy, $\lambda$ is bounded from below according to  (\ref{entrocon}). This implies an upper bound on $\alpha$ given by $8\sqrt{3}$.
In FIG. 1 we observe that conductivity approaches to a constant value when increasing $\alpha$. Its minimum value is obtained asymptotically approaching to $\alpha=2\sqrt{3}$ from the right.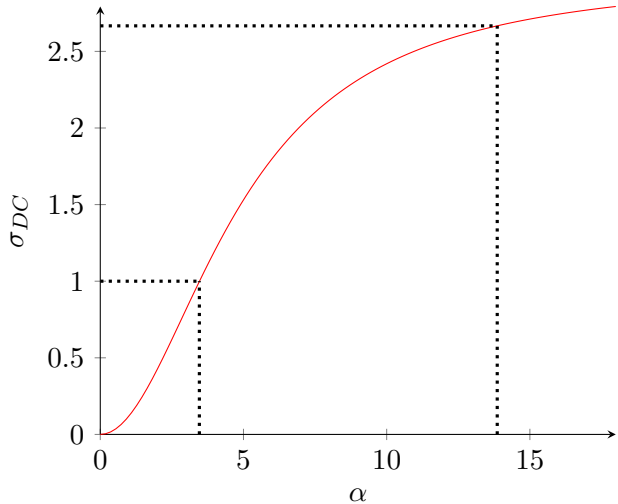
\begin{figure}[H]
\centering
\begin{tikzpicture}
\begin{axis}[
    axis lines = left,
    xlabel = $\alpha$,
    ylabel = {$\sigma_{DC}$},
]
\addplot [
    domain=0:18, 
    samples=100, 
    color=red,
]
{3*x^2/(24+x^2)};
%\addlegendentry{$$}
\draw [very thick,dotted] (3.46,0) -- (3.46,1);
\draw [very thick,dotted] (0,1) -- (3.46,1);
\draw [very thick,dotted] (13.86,0) -- (13.86,8/3);
\draw [very thick,dotted] (0,8/3) -- (13.86,8/3);
\end{axis}
\end{tikzpicture}\caption{DC conductivity respect to $\alpha$. Between the dashed lines, this curve represents conductivities for physically acceptable configurations.}
\end{figure}
 In addition, through the relation \eqref{AdScriterium} and \eqref{temp}, it can be seen that the Hall angle decays as $\theta_H\sim 1/T^2$ at high temperature limit, resembling the quadratic-T inverse behavior found in cuprates holographically studied in \cite{Blake:2014yla,Jiang:2017imk,Ge:2016sel}.

To obtain a richer phenomenology, we have to modify the axionic part of the action by means of nonminimal couplings. Because our axion depends linearly on the flat coordinates, we need to modify the action in such a way that maintains the original shift symmetry, otherwise the transverse manifold coordinates would appear on the equations of motion. The most simple case to do this is to consider a k-essence term as we did it in \cite{Cisterna:2017jmv}

\begin{eqnarray}\label{actionnew}
\bar{S}[g,\phi,\psi_I]=\int{d^4x\sqrt{-g}\left[R-2\Lambda-\frac{(\partial\phi)^2}{2}-\frac{\phi^2 R}{12}-\sum_{I=1}^{2}\left(\frac{X_I}{2}+\beta\left(\frac{X_I}{2}\right)^k\right)-\frac{F^2}{4}\right]}
\end{eqnarray}
where we have defined $X_I:=\partial_{\mu}\psi_I\partial^{\mu}\psi_I$.
This goes back to the action \eqref{action} plus \eqref{actionMax}  when $\beta$, with mass dimension $4-4k$, goes to zero. The equations of motion for the above action are shown in the Appendix B.
We know that when $\phi(r)=0$, the case $k=2$ modifies the black brane solution found in \cite{andrade} including a term which goes as $1/r^2$, like the electric charge. Then we provides an exact solution for (\ref{actionnew}) at this particular case
\begin{eqnarray}
ds^2&=&-\frac{(r-3\lambda l)(r+\lambda l)^3}{r^2 l^2}dt^2+\frac{r^2 l^2}{(r-3\lambda l)(r+\lambda l)^3}dr^2+r^2(dx^2+dy^2), \label{soluNew}\\
\psi_I&=&2\sqrt{3}\lambda x^I\label{axionfieldNew},
\end{eqnarray}
and the scalar field and the Maxwell gauge field read
\begin{align}
\phi&=\frac{\sqrt{Q_E^2+12\lambda^4 l^2+576\lambda^4 \beta}}{\lambda(r+\lambda l)},\ &\ A&=-\frac{Q_E}{r}dt. \label{sgfields2}
\end{align}
Here we are interested in the DC conductivity, so we turn off the magnetic field. For this solution, the AdS criterium is given by
\begin{equation}\label{AdScriterium2}
\lambda^4=\frac{Q_E^2}{(\alpha^{2}-12)l^2-576\beta},
\end{equation}
requiring
\begin{equation}\label{condition}
(\alpha^{2}-12)l^2>576\beta.
\end{equation}
With the same method and algebra computation as the previous study, we obtain the DC  conductivity
\begin{eqnarray}
\bar{\sigma}_{DC}&=&\frac{3 \left[12 (48 \beta +1) \lambda ^4+Q_E^2\right]}{36 (48 \beta +1) \lambda^4+Q_E^2}=1+\frac{2Q_E^2}{36(48 \beta +1)\lambda ^4+Q_E^2}\nonumber\\
&=&1+\frac{2(\alpha^2-12-576\beta)}{\alpha^2+24+1152\beta}=\frac{3\alpha^2}{\alpha^2+24+1152\beta}    \label{SCK}
\end{eqnarray}
where we have considered \eqref{AdScriterium2} with $l=1$ in the second line.\\
The above result is also independent of $\lambda$ and it reduces to the DC conductivity \eqref{conduc} when $\beta=0$ as we expected. We will consider only positive values of $\beta$ in order to avoid phantom contributions into the axionic sector. It is observed that the conductivity depends on two parameters, $\alpha$ and $\beta$ which satisfy \eqref{condition}, constraining $\beta$ to take values in the interval $[0,\frac{\alpha^2-12}{576}[$, otherwise the axionic parameter may become complex. In this case, $\alpha$ is still bounded from above by requiring configurations of positive entropy (\ref{entrocon}) such that $\alpha<8\sqrt{3}$. For these considerations we note that in the region of parameters of physically acceptable configurations, the conductivity again reaches its maximum value $\sigma_{DC}\rightarrow 8/3$ for $\alpha\rightarrow 8\sqrt{3}$ and $\beta=0$. Now, the inclusion of the $\beta$ parameter by means of the k-essence contribution, allows to access to a wider range of conductivities. Namely, all the configurations for which $\beta$ approaches its upper bound, it is $\beta\rightarrow\frac{\alpha^2-12}{576}$, possesses $\sigma_{DC}\rightarrow 1$. This means that the conductivity now lies on a two dimensional surface described by $\alpha$ and $\beta$, whose extremal values gives $\sigma_{DC}\in ]1,8/3[$. We have to note that (\ref{condition}) prevents our solution to behaves as an insulator, as it imposes a lower bound for $\beta$ that forbids to access the insulator state for physically acceptable configurations. Fig. 2 shows this particular behavior.

\begin{figure}[H]
\centering
\includegraphics[width=14cm]{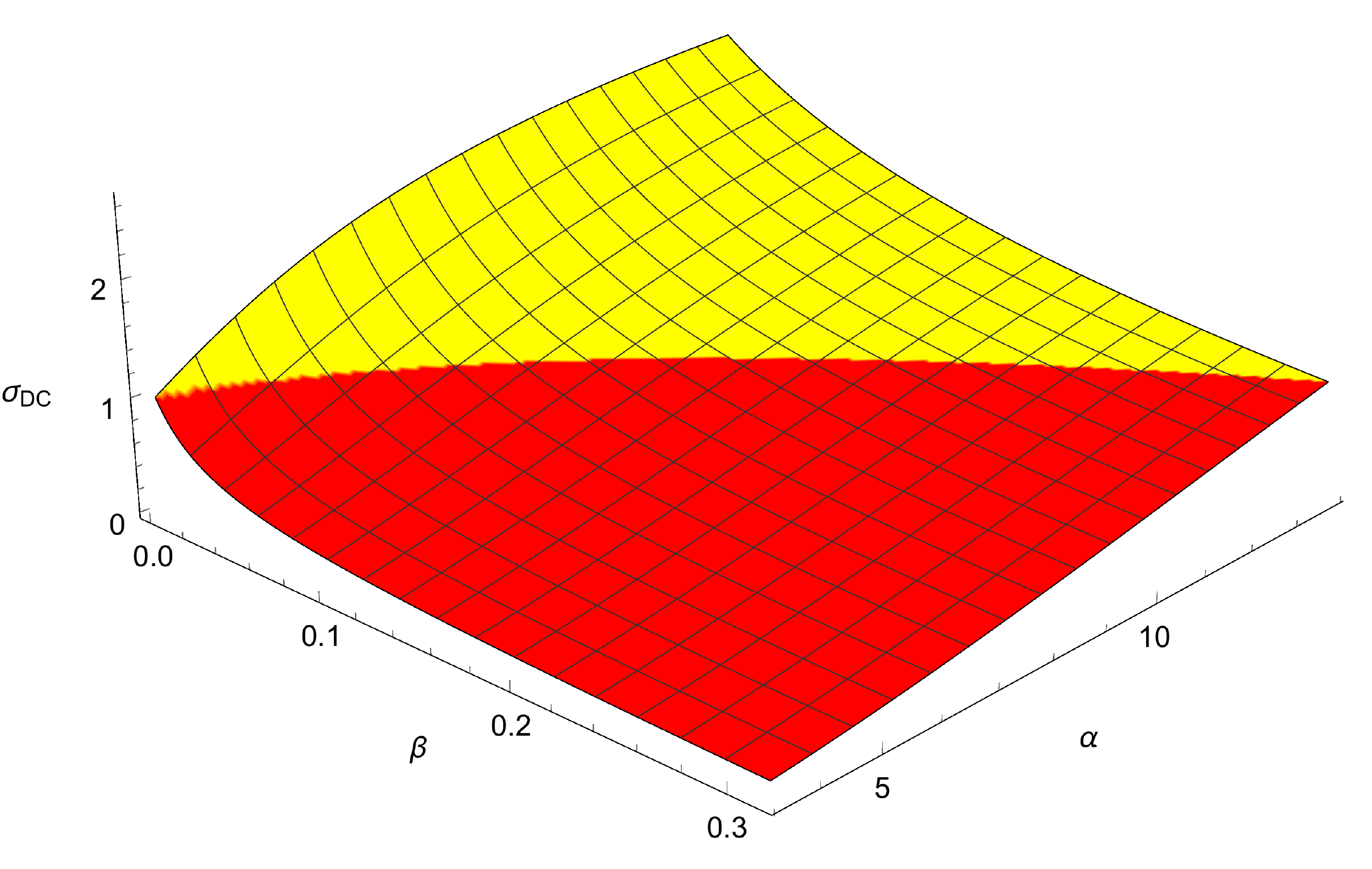}
\caption{DC conductivity respect to $\alpha$ and $\beta$. The yellow region represents conductivities for physically acceptable configurations.}
\end{figure}
\section{Further remarks}

In this paper we have constructed black brane solutions in a conformally coupled scalar theory. These solutions are supported by two axion fields homogeneously distributed along the horizon that depend linearly on the transverse manifold coordinates. No self-interaction for the scalar field is needed. Solutions of this kind have been shown to produce momentum dissipation making them excellent candidates to study holographic conductivities \cite{andrade}.
We observe that, for the neutral case, all parameters appearing in the solution are free from fine tuning. The electric and magnetically charged case is also obtained. It is shown in this case that, in order to the asymptotic behavior of the scalar field to respect conformal invariance, the axionic and electromagnetic charges must be related by means of \eqref{AdScriterium}. This implies that no contribution from the scalar fields appears on the mass. We then perform in detail the thermodynamic analysis of our charged solutions considering the non-trivial effect of the axionic charges. The rotating black brane can be obtained from the static one, by performing a boost in the plane spanned by the temporal coordinate and one of the planar coordinate. It has been shown that the physically acceptable configurations with finite conserved charges, possesses mass and angular momentum tuned by the axionic parameter. This configuration satisfies the dominant energy condition and is thermodynamically locally stable, which means that always attains equilibrium with a heat bath as well as its static counterpart.\\
Following the procedure stated in \cite{q2}, and using the momentum relaxation techniques of \cite{andrade}, we obtain the DC conductivity and Hall angle of our charged solutions. We demonstrated that due to our AdS criterium \eqref{AdScriterium} our solution mostly behaves as the neutral case originally studied in \cite{q2}. Nevertheless, in this case we have $\sigma_{DC}>1$ and, contrary to the charged black branes originally constructed in \cite{Bardoux:2012aw} and holographically studied in \cite{andrade}, its behavior is totally independent from the temperature. Despite this, our DC conductivity depends on the arbitrary constant $\alpha$ that ensure a non trivial scalar field $\phi(r)$, whose asymptotic behavior respects the required conformal symmetry. It is found that in the range of values for this constant that ensures physically acceptable configurations, the conductivity profile is a monotonically increasing function of $\alpha$, taking  values from $\sigma_{DC}=1$ to $\sigma_{DC}=8/3$. In order to look for different behaviors of our holographic conductor we have included a new contribution in the axionic sector of our system. Motivated by the result obtained in \cite{Cisterna:2017jmv} for k-essence theories, we study the k-essence case with $k=2$. This case is integrated due to the fact that the k-essence contribution into the equations of motion possesses the same behavior as the electric charge. We observe that our AdS criterium is modified but the DC conductivity is still independent from the temperature. This behavior was expected since the $k=2$ term behaves similarly to the electric charge. Despite this, the new coupling $\beta$ modifies the conductivity in such a way that, while respecting the condition (\ref{condition}), $\sigma_{DC}\sim 1$ for a wide range of values for $\beta$ given by $\beta\rightarrow\frac{\alpha^2-12}{576}$. We observe that, for small values of $\alpha$ and large values of $\beta$, $\sigma_{DC}\sim 0$ behaves as an insulator. Nevertheless, this region belongs to unphysical configurations since they violates the reality condition for the axion fields. As a final remark, we can conclude that, either in the case of pure minimally coupled axion fields or in more exotic theories as in k-essence, conformally scalar fields that respect the asymptotic AdS symmetry, renders the DC conductivity to a constant value that depends on the coupling parameters and not on the temperature as it is usually expected.\\
As further development of this work, we will disclose the thermal conductivity, thermoelectric conductivity of the dual boundary theory elsewhere. Moreover, we expect to study more general conformally coupled scalar theories, as the one studied in \cite{Oliva:2011np}, that exists in dimension higher than four. These would not only provide black brane solutions for such models but also  a new setup to study holographic conductivities and their phenomenology. %\red{From the viewpoint of experiment, we expect the features of conductivities in our model can be observed in a real laboratory in the future.}

\section*{ACKNOWLEDGMENTS}
A. C would like to thank Luciano Vanzo and Mokhtar Hassaine for interesting comments and discussions. A.C. work is supported by FONDECYT Grant N\textsuperscript{o}11170274 and Proyecto Interno Ucen I$+$D-$2016$, CIP2016. C.E. acknowledges financial support given by Becas Chile, CONICYT. X. M. Kuang is supported by the Natural Science Foundation of China under Grant No. 11705161 and Natural Science Foundation of Jiangsu Province under Grant No. BK20170481.

\section{APPENDIX}
\subsection{Rotating black branes}  \label{secapenA}
When the static solution possesses a planar base manifold, a rotating solution can be constructed from the static one by applying an improper local transformation \cite{Erices:2015xua}. Additionally, as it was noted in \cite{Stachel}, by performing a topological identification along one of the planar coordinates, it is possible to obtain a stationary solution provided with angular momentum. For simplicity, we will consider the non charged solution. To achieve this, the $x$ coordinate now dubbed as $\varphi$ and ranging $\varphi\in(-\infty,\infty)$, along with the $t$ coordinate are transformed as follows,
\begin{align}
t &\rightarrow \frac{1}{\sqrt{1-\omega^2}}(t-l\omega \varphi)\ ,&\ \varphi &\rightarrow \frac{1}{\sqrt{1-\omega^2}}\left(\varphi-\frac{\omega}{l} t\right).   \label{improper}
\end{align}
This transformation is a boost in the $t-\varphi$ plane parametrized by $\omega^2<1$. By means of a topological identification, the initial topology of the base manifold $\mathbb{R}\times\mathbb{R}$ can be transformed to $S^1\times\mathbb{R}$, by fixing the period of the angular coordinate $\varphi$ to $2\pi$. This implies that the local geometry of the static solution is preserved but not the global one. Hence, the transformation \eqref{improper} provides a resulting manifold which is globally stationary but locally static.\\
Applying (\ref{improper}) onto our original static solution (\ref{solu})-(\ref{axionfield}) we obtain 
\begin{equation}\label{rotating}
ds^2=-N^2(r) F(r) dt^2+\frac{dr^2}{F(r)}+H(r)(d\varphi+N^{\varphi}(r)dt)^2+r^2 dy^2
\end{equation}
where $F(r)$ is given by (\ref{solu}) and 
\begin{align}
N^2(r)&=\frac{r^2(1-\omega^2)}{r^2-l^2\omega^2 F(r)},\ & \  N^{\varphi}(r)&=-\frac{r^2-l^2F(r)}{r^2-l^2\omega^2 F(r)}\frac{\omega}{l},\ & \  H(r)&=\frac{r^2-l^2\omega^2F(r)}{1-\omega^2}. 
\end{align}
On the other hand the conformally coupled scalar field $\phi(r)$ is not affected by the transformation while the axion field originally distributed along the $x$ direction becomes time dependent 
\begin{equation}
\psi_1=2\sqrt{3}\frac{\lambda}{\sqrt{1-\omega^2}}\left(\varphi-\frac{\omega}{l}t\right).
\end{equation}
From \eqref{rotating} we see that that the rotating solutions still possesses a single event horizon located at $r_+=3\lambda l$ overing the curvature singularity in $r=0$. The functions $N^2(r)$ and $H(r)$ are both positive, which ensures the proper signature for the black hole and a well defined local area for the base manifold. All of these functions are monotonically increasing functions outside horizon. The asymptotic behavior corresponds to a boosted AdS spacetime.\\
It is possible to show that this solution indeed describes a rotating black hole. For this, we follow the Regge-Teitelboim approach \cite{ReggeTeitelboim} to determine the mass and angular momentum for this rotating solution. The Hamiltonian generator of the asymptotic symmetries $\xi^{\mu}=(\xi^{\perp},\xi^{i})$ for the Lagrangian in \eqref{action}, is given by a linear combination of the Hamiltonian constraints $\mathcal{H}_{\perp}, \mathcal{H}_i$ supplemented with a surface term $Q[\xi^{\mu}]$ which ensures well-defined functional derivatives for the Hamiltonian generator. In the ADM decomposition, $\gamma_{ij}$ is the metric of the spacelike surfaces of constant time. Along with the scalar field and the axion fields $\psi_{I}$, they constitute the dynamical variables of the system with conjugate momenta $\pi_{ij}$, $\pi_{\phi}$ and $\pi_{\psi_I}$, respectively. The generator reads,
\begin{equation}\label{generator}
H[\xi^{\mu}]=\int d^{3} x \left(  \xi^{\perp}
\mathcal{H}_{\perp}+\xi^{i}\mathcal{H}_{i}\right) +Q[\xi^{\mu}].
\end{equation}
For the kind of configurations considered \eqref{rotating}, the expressions for the constraints are explicitly given by \cite{EricesMartinez02}
\begin{align}
\mathcal{H_{\bot}}=&-\sqrt{\frac{H}{F}}r\left[\left(\kappa-\frac{\xi}{2}\phi^2\right){^{(3)}R}-\frac{1}{2} F (\partial_r\phi)^{2}-\kappa \Lambda\right] \nonumber \\&+\sqrt{\frac{f}{H}}\left(\frac{4}{2\kappa-\xi\phi^2}\right)\pi_{r\varphi}\pi^{r\varphi}+\sqrt{\frac{f}{H}}\frac{\pi_{\psi_1}^2}{2r}-\xi\partial_r( \sqrt{f H}r\partial_r\phi^{2}),  \label{hperp}\\
\mathcal{H}_{\varphi}=&-2{{\pi_{\varphi}}^{r}}_{|r}+\pi_{\psi_1}\partial_{\varphi}\pi_{\psi_1}\, \label{Hphi}
 \end{align} 
where ${}^{(3)}R$ is the Ricci scalar of $\gamma{ij}$. The nonvanishing components of the momenta are given by,
\begin{align}
\pi_{\varphi}^{\ r}&=-\frac{rH^{3/2}}{2N}\left(\kappa-\frac{\xi}{2}\phi^2\right) \partial_r N^{\varphi},&\pi_{\psi_1}=\frac{r\sqrt{H}}{N F}(\partial_r\psi_1-N^{\varphi}\partial_{\varphi}\psi_1). 
\end{align}
Demanding $\delta H=0$ on-shell, we obtain the variation of the charge
\begin{align} 
&\delta Q[\xi^{\mu}]=\lim_{r \to \infty} \left\{\sigma \sqrt{\frac{H}{F}}\left[\left(-\frac{F}{H}\left(\partial_r \delta H-\frac{\partial_r H}{2H}\delta H \right) 
-\left(\frac{\partial_r H}{2H}+\frac{1}{r}\right)\delta F\right)\left(\frac{1-\kappa\xi\Phi^2}{2\kappa}\right)\xi^{\bot} \right.  \right. \nonumber \\
&\left. \left.+\frac{F}{H}\partial_r\left[ \left(\kappa-\frac{\xi}{2}\phi^2\right)\xi^{\bot}\right]\delta H+F\left[ \xi^{\bot}(\xi\partial_r (\delta\phi^2)-\delta\phi \partial_r \phi)-\xi\partial_r \xi^{\bot}\delta\phi^2 \right] \right]+2\sigma\xi^{\varphi}\delta\pi_{\varphi}^{\ r}\right.
\nonumber \\ & \left.-\int dS_i \left[\frac{r\sqrt{H}}{F}\xi^{\perp}\partial^{i}(\phi\delta\phi+\psi_1\delta\psi_1+\psi_2\delta\psi_2)+\xi^i\pi_{\psi_1}\delta\psi_1\right]\right\}, \label{dq}
\end{align}
where $\sigma$ stands for the volume of the base manifold.
The deformation vectors $\xi^{\mu}$ in terms of the Killing vectors $\partial_t$ and $\partial_{\varphi}$ are given by,
\begin{align}
\xi^{\bot}&=N\sqrt{F}  \partial_t,\\
\xi^{\varphi}&=\partial_\varphi+N^{\varphi}  \partial_t.
\end{align}
The mass $M$ is the conserved charge associated with the time translation symmetry, while the angular momentum $J$ is the conserved charge associated with the rotatonal translation symmetry. In fact, in this approach, both of them are obtained by evaluating $\delta \mathcal{M}=\delta Q[\partial_t]$ and $\delta \mathcal{J}=-\delta Q[\partial_\varphi]$, respectively. Namely,
\begin{eqnarray}
\delta \mathcal{M}&=&\lim_{r \to \infty} \left\{(t+\beta_\varphi l\omega)\delta\Omega\frac{r}{l}+\frac{24l\sigma\lambda^2}{(1-\omega^2)^2}\left[(\omega^4+\omega^2-2)\delta\lambda-2\lambda\omega\delta\omega\right]\right\}, \label{m}\\
\delta \mathcal{J}&=&\lim_{r \to \infty} \left\{-(\omega t+\beta_\varphi l)\delta\Omega r +\frac{24l^2\sigma\lambda^2}{(1-\omega^2)^2}\left[3(1-\omega^2)\omega\delta\lambda+(1+\omega^2)\lambda\delta\omega\right]\right\} \label{j}
\end{eqnarray}
where $\delta\Omega$ is given by,
\begin{equation}
\delta\Omega=- \frac{12\lambda\beta_y}{(1-\omega^2)^2}\left[\omega(1-\omega^2)\delta\lambda+\lambda\delta\omega\right]\ .
\end{equation}
As it can be seen in \eqref{m} and \eqref{j}, the boost performed to the static solution, has introduced those terms proportional to $\delta\Omega$ which are divergent. However, we can obtain finite conserved charges by demanding $\delta\Omega=0$, which imposes a relation between the boost parameter $\omega$ and the axion parameter $\lambda$. This is determined by a differential equation of the form
\begin{equation}
\frac{d\omega}{d\lambda}+\frac{\omega}{\lambda}(1-\omega^2)=0,
\end{equation}
whose solution gives
\begin{equation}
\omega=\frac{\omega_0}{\sqrt{\omega_0^2+\lambda^2}}\ ,
\end{equation}
where $\omega_0$ is a constant without variation. This relation renders the variation of the charges finite and the remaining terms can be directly integrated. Therefore, the mass and momentum angular have the following expressions,
\begin{align}
\mathcal{M}&=8\sigma l \lambda(3\omega_0^2+2\lambda^2),&\mathcal{J}&=24\sigma l^2 \lambda\omega_0\sqrt{\omega_0^2+\lambda^2},
\end{align}
up to additive fixed constants. This constants have been fixed in order to have a static and massless background. In other words, switching off the axion field through $\lambda=0$, the configuration
\begin{equation} \label{bg}
ds^2=-\frac{r^2}{l^2}dt^2+\frac{l^2}{r^2}d r^2+r^2 (d\varphi^2+dy^2),
\end{equation}
possesses $\mathcal{M}=0$ and $\mathcal{J}=0$. Note also that, naturally the angular momentum vanishes when the boost parameter vanishes, recovering the mass for the static solution obtained in Sec. \ref{secthermo} in the neutral case. Also, it is worth to note, that the angular momentum is bounded from above by the mass $|\mathcal{J}|<\mathcal{M} l$.\\
The temperature of the black hole can be computed by means of the surface gravity $k^2=-\frac{1}{2}\nabla_\mu\chi_\nu\nabla^\mu\chi^\nu$ given in terms of the null Killing vector at the event horizon $\chi=\partial_t+\Omega_+\partial_\varphi$. Here, $\Omega_+=\omega/l$ is the angular velocity at the horizon. This is
\begin{equation} \label{temperature}
T=\frac{k}{2 \pi}=\frac{16}{9\pi} \frac{\lambda^2}{l\sqrt{\omega_0^2+\lambda^2}},
\end{equation}
which, as it was expected, reduces to \eqref{temp} when $\omega_0=0$. The conformal coupling modifies the standard Bekenstein-Hawking entropy, giving \cite{Visser, Ashtekar}
\begin{equation}
S=\frac{A_+}{4\tilde{G}_+},
\end{equation}
where $A_+=\sigma r_+\sqrt{H(r_+)}$ is the area of the horizon. As it was expected, the first law of thermodynamics is satisfied,
\begin{equation}
d\mathcal{M=}Td\mathcal{S}+\Omega_+d\mathcal{J}+\Phi_{\psi_1} d\mathcal{Q}_1+\Phi_{\psi_2} d\mathcal{Q}_2,
\end{equation}
with chemical potentials for axion fields and axionic charges determined by
\begin{align}
\Phi_{\psi_1}&=2\sqrt{3(1-\omega^2)}\lambda r_+,&\mathcal{Q}_{\psi_1}&=-\frac{2\sqrt{3}\lambda\sigma}{\sqrt{1-\omega^2}},\\
\Phi_{\psi_2}&=2\sqrt{3\lambda} r_+,&\mathcal{Q}_{\psi_2}&=-2\sqrt{3}\lambda\sigma.
\end{align}
The local thermal stability of the black hole can be analyzed by computing the specific heat at fixed angular velocity and chemical potentials,
\begin{equation}
C_{\Omega_+, \Phi_{\psi_I}}=\left(\frac{\partial M}{\partial T}\right)_{\omega, \Phi_{\psi_I}}=\left[\left(\frac{\partial M}{\partial r_+}\right)\left(\frac{\partial T}{\partial r_+}\right)^{-1}\right]_{\omega, \Phi_{\psi_I}},
\end{equation}
which gives
\begin{equation}
C_{\Omega_+, \Phi_{\psi_I}}=\frac{3\pi\sigma(r_+^2+9l^2\omega_0^2)^{3/2}(2 r_+^2+9l^2\omega_0^2)}{2 r_+(r_+^2+18l^2\omega_0^2)}.
\end{equation}
The specific heat  is always positive, and in consequence, the rotating black hole always attains equilibrium with a heat bath as well as the static one ($\omega_0=0$).

\subsection{Equations of motion for the k-essence extension}
Einstein equations for \eqref{actionnew} are given by
\begin{eqnarray}
\kappa(G_{\mu\nu}+\Lambda g_{\mu\nu})=\frac{1}{2} T^{\phi}_{\mu\nu}+\frac{1}{2}T^{\psi}_{\mu\nu}+\frac{1}{2}T^{em}_{\mu\nu}
\end{eqnarray}
where we have defined
\begin{eqnarray}
T^{\phi}_{\mu\nu}&=&\partial_\mu\phi\partial_\nu\phi-\frac{1}{2}g_{\mu\nu}(\partial\phi)^2+\frac{1}{6}(g_{\mu\nu}\Box-\nabla_\mu \nabla_\nu+G_{\mu\nu})\phi^2,\\
T^{\psi}_{\mu\nu}&=&\sum_{I=1}^2\left[\partial_{\mu}\psi_I\partial_{\nu}\psi_I-\frac 12 g_{\mu\nu}X_I+2\beta(kX_I^{k-1}\partial_{\mu}\psi_I\partial_{\nu}\psi_I-\frac 12 g_{\mu\nu}X_I^k)\right]\\
T_{em}&=&F_{\mu\rho}F_{\nu}^{\rho}-\frac{1}{4}g_{\mu\nu}F_{\alpha\beta}F^{\alpha\beta}.
\end{eqnarray}
On the other hand, for our conformally coupled scalar and axion fields we have
\begin{eqnarray}
\left(\Box-\frac{1}{6}R\right)\phi=0,\\
\left[(1+\beta kX_I^{k-1})g^{\mu\nu}+\beta
k(k-1)X_I^{k-2}\nabla^{\mu}\psi_I\nabla^{\nu}\psi_I\right]\nabla_{\mu}\nabla_{\nu}\psi_I=0.
\end{eqnarray}
These equations provides our solution (\ref{soluNew}), (\ref{axionfieldNew}) and (\ref{sgfields2}), for the $k=2$ case.

\end{document}